%% file: main.tex
\documentclass[conference]{IEEEtran}
\IEEEoverridecommandlockouts
\PassOptionsToPackage{bookmarks=false}{hyperref}
\renewcommand{\figurename}{Figure}

\ifCLASSINFOpdf
  \usepackage[pdftex]{graphicx}
\else
\fi

\usepackage[latin1,utf8]{inputenx}
\usepackage[cmex10]{amsmath}
\usepackage{algorithmic}
\usepackage{algorithm}
\usepackage{array}
\usepackage{xcolor}
\usepackage{hyperref}
\hypersetup{colorlinks=true,citecolor=blue,linkcolor=blue,urlcolor=blue,plainpages=false}
\usepackage{url}
\usepackage{booktabs, tabularx, multirow, array}
\usepackage{nomencl}
\usepackage{comment} 
\usepackage{amsmath}
\usepackage{eurosym}
\usepackage{cite}

\usepackage{array, amsthm, amsmath ,amssymb, amsfonts, booktabs,cancel, comment,enumitem,eurosym,float,graphicx,mathtools, multirow,placeins,soul,steinmetz,stfloats,siunitx,subfigure,svg,url,xcolor,rotating}

\usepackage{etoolbox}
\renewcommand\nomgroup[1]{%
  \item[\bfseries
  \ifstrequal{#1}{I}{\textit{Indices and Sets}}{%
  \ifstrequal{#1}{P}{\textit{Parameters}}{%
  \ifstrequal{#1}{F}{\textit{Functions}}{%
  \ifstrequal{#1}{M}{\textit{Production Models}}{%
  \ifstrequal{#1}{V}{\textit{Variables}
  }{}}}}}%
]}

\floatstyle{ruled}
\newfloat{model}{thp}{lop}
\floatname{model}{\textsc{Model}}

\DeclareSIUnit\h{h}
\DeclareSIUnit\A{A}
\DeclareSIUnit\Ah{Ah}
\DeclareSIUnit\MWh{MWh}
\DeclareSIUnit\kWh{kWh}
\DeclareSIUnit\MW{MW}
\DeclareSIUnit\MVAr{MVAr}
\DeclareSIUnit\MVA{MVA}
\DeclareSIUnit\pu{p.u.}
\DeclareSIUnit\EurokWh{\text{\euro}/kWh}
\DeclareSIUnit\EuroMWh{\text{\euro}/MWh}
\usepackage{cite}
\usepackage{algorithmic}
\usepackage{graphicx}
\usepackage{textcomp}
\usepackage{xcolor}
\def\BibTeX{{\rm B\kern-.05em{\sc i\kern-.025em b}\kern-.08em
    T\kern-.1667em\lower.7ex\hbox{E}\kern-.125emX}}

\newcommand{\T}{\text{tech}}

\makenomenclature
\begin{document}

\title{Empirical Analysis of Capacity Investment Solution in Distribution Grids}

\author{\IEEEauthorblockN{Luis Lopez, Alvaro Gonzalez-Castellanos, David Pozo}
\IEEEauthorblockA{Center for Energy Science and Technology \\
Skolkovo Institute of Science and Technology (\textit{Skoltech}) \\
Moscow, Russia \\
\{luis.lopez, alvaro.gonzalez, d.pozo\}@skoltech.ru}}

\maketitle
\begin{abstract}
This paper presents an analysis of the stability and quality of the distributed generation planning problem's investment solution. The entry of distributed generators power based on non-conventional energy sources has been extensively promoted in distribution grids. In this paper, a two-stage stochastic programming model is used to find the optimal distributed generators' installed capacities. We emphasize the design of scenarios to represent the stochasticity of power production on renewable sources. In the scenario generation, a method is proposed based on the clustering of real measurements of meteorological variables. We measure the quality and stability of the investment solution as a function of the number of scenarios. The results show that a reduced selection of scenarios can give an inadequate solution to distributed generators' investment strategy. 
\end{abstract}

\printnomenclature[0.55in]

\IEEEpeerreviewmaketitle
\section{Introduction} \label{Sec:Intro}

In modern distribution networks, users can inject active power into the grid by small power plants \cite{Kakran2018SmartReview}. The power plants connected near the demand buses are called distributed generation units (DG units) \cite{Singh2017APlanning}. Large amounts of power are being injected through DG units due to policies that promote non-conventional renewable energies in different countries \cite{PereiradaSilva2019PhotovoltaicAdaptation, Duong2017TheSystem, Dong2016CleanExperience}. With DG's widespread deployment, the distribution system operator needs to plan and coordinate the new DG units' installation capacity. DG planning can reduce operating costs or solve technical restrictions \cite{Jain2017DistributedEra}.

The investment solution in distribution networks refers to determine the installed capacities and locations of DG units. When DG units are power-based on non-conventional renewable technologies, they behave as non-controllable and stochastic negative load. Thus, we need to capture the uncertainty associated with meteorological measurements \cite{Notton2018IntermittentForecasting}. Modeling the stochasticity of renewable generation sources has been widely confronted by several authors \cite{Zakaria2020UncertaintyApplications, Gonzalez-Castellanos2019StochasticStorage, Singh2017APlanning, Jain2017DistributedEra}. Jooshaki \textit{et al.} \cite{Jooshaki2020AUnits} propose a tool to integrate DG units using a mixed-integer linear stochastic model and perform a case study on a 24-node distribution network. In \cite{Home-Ortiz2019AMitigation}, the authors proposed a methodology using mixed-integer stochastic programming to find the best reinforcement plan for mitigating greenhouse gas emissions. In \cite{DeLima2019AEmissions}, a stochastic model is proposed to address the problem of distribution system expansion with uncertainties of DG units and issues related to CO2 emissions \cite{Lopez2015APlanning}. 

Stochastic programming is a mathematical framework that lets capturing the uncertainty of power production from non-conventional renewable sources \cite{Narayan2017Risk-averseUncertainty, Zhou2013ASystems}. It has been proposed in \cite{Seljom2019SampleDesign} to use Sample Average Approximation (SAA) to generate scenarios in the planning problem with stochastic parameters. Nevertheless, scenario generation techniques are limited because they are an approximation (discrete scenarios) of real distribution. Therefore, the stochastic model relies on scenario representation, and if scenario representation is deficient, information about the actual probability distribution may be lost.  This work proposes a two-stage stochastic programming model that provides an investment solution considering short-term uncertainty in a long-term planning problem. We propose the k-means clustering technique for the scenario generation to reduce the problem's dimensionality and capture the underlying correlation between the random variables. We analyze the quality of the investment solution as a function of the number of scenarios used. We calculate the expected value and the dispersion of the solution obtained and upper and lower bound of the investment solution for each number of scenarios. 

The main contribution of this work is the analysis of the quality and stability of the investment solution in the DG planning problem using empirical measurements. We assess how the investment solution deviates from its ground-truth value when we use an inadequate description of the problem's stochasticity (few numbers of scenarios). The work is organized as follows: Section \ref{Sec:Methods} describes the two-stochastic programming model and the estimation of the upper and lower boundaries. Section \ref{Sec:Test} introduces the case study and the scenario generation technique. Section \ref{Sec:Results} shows the findings and simulations performed on a test distribution system with real measurements. Section \ref{Sec:Conclusions} provides the discussions and conclusions of the observed empirical stability. 
\section{Methodology} \label{Sec:Methods}

Stochastic programming provides solutions using scenarios to represent possible realizations of the uncertainty. This section describes our methodology for solving the problem of investment in DG units using stochastic programming. In Section \ref{Sec:Two_Stage}, we briefly describe the model used and in Section \ref{Sec:Stability}, we describe the metrics to evaluate the quality of the solution obtained.  

\subsection{Two-stage problem formulation} \label{Sec:Two_Stage}

\begin{figure}[htbp]
    \centering
    \includegraphics[width=\columnwidth]{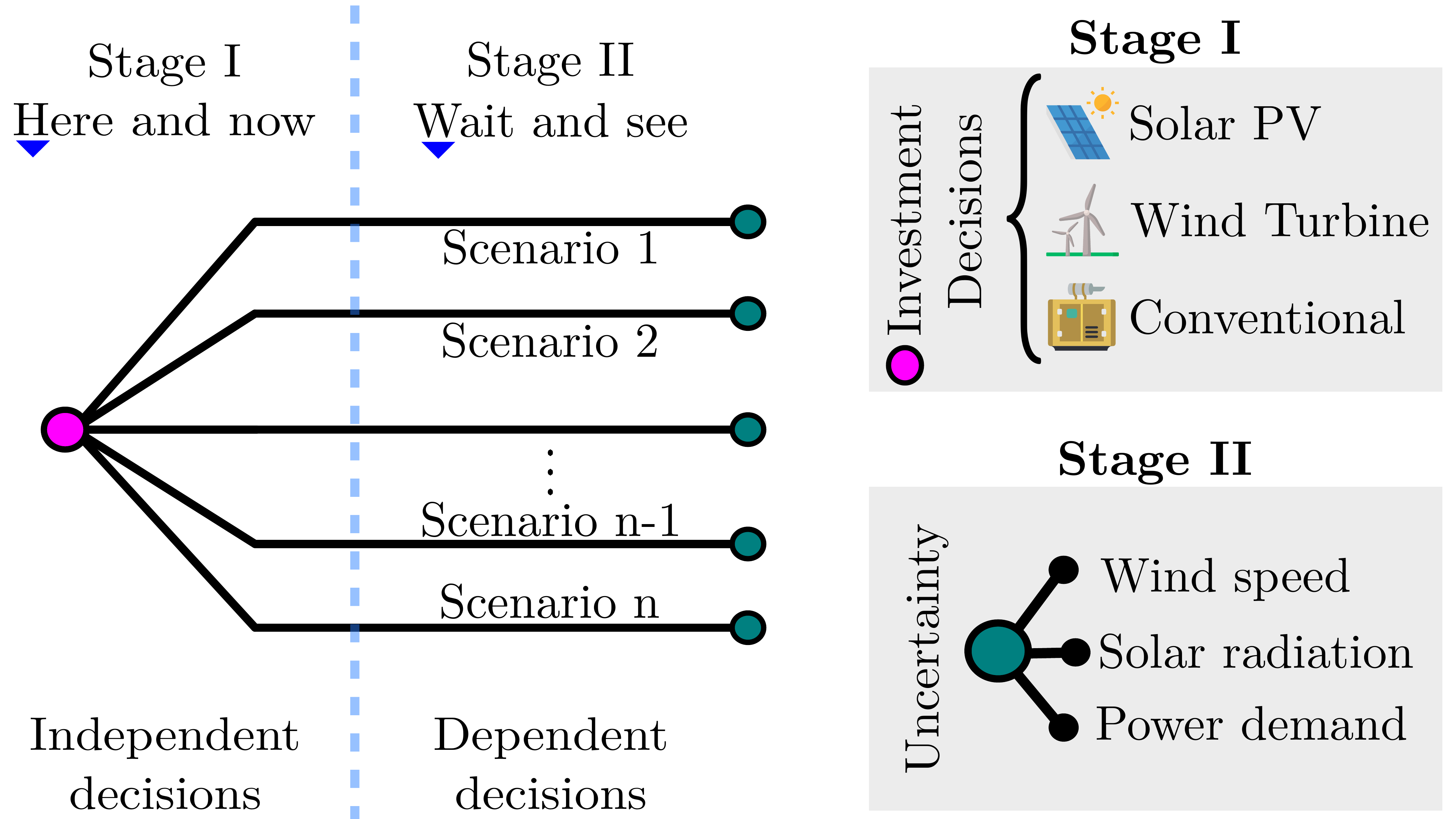}
    \caption{Two-stage stochastic approach.}
    \label{fig:Two_stage}
\end{figure}

This article addresses the problem of DG planning through a two-stage stochastic programming approach (Fig. \ref{fig:Two_stage}). The first stage consists of the investment solution of the DG units. Three technologies of DG units are considered: solar photovoltaic (PV), wind turbines (WT), and conventional generators (CG). The first-stage variables are integers since the power plant units are manufactured in discrete modules of installed power. The second stage consists of the computation of the operation and maintenance cost for every scenario. The second stage calculates the expected value of the power produced by the newly installed DG units. The uncertainty of power production and energy balance is associated with the meteorological variables of solar radiation, wind speed, temperature, and energy demand. The evaluation of the expected value of power production given an investment decision requires numerous second-stage optimization problems that depend on the number of scenarios. In summary, this stochastic DG planning problem involves a large number of scenarios and integer variables in the first stage. 

\input{formulation.tex}

The objective function (\ref{Eq: of}) minimizes the system's energy cost over the analyzed time horizon. The energy cost is the sum of the investment cost and the total cost of operation and maintenance. The investment cost (\ref{Eq: investment}) is equal to the sum of the installation costs per technology in each node.  The total cost of operation and maintenance (\ref{Eq: expected_value_prob}) is the sum of the probability of occurrence multiplied by operation and maintenance costs in each of the scenarios generated. Each scenario's operation and maintenance costs depend on the costs of active power losses, the energy imported from the power grid, and the new DG units' power production costs. Demand profiles and weather conditions only depend on the scenario and not on the system nodes since distribution networks cover the same area. 

The constraints of the stochastic DG planning approach are divided into physical law constraints (\ref{Eq: active_powerflow})-(\ref{Eq: current_flow}) and engineering constraints (\ref{Eq: volt_limit})-(\ref{Eq: pf_lead_eq}). The physical law constraints are the active and reactive power balance constraints (generated power must supply the demand power) and the power flow constraints through the transmission lines. The power flow constraints are modeled through the DistFlow equations \cite{Baran1989OptimalSystem, Baran1989OptimalSystems}. The McCormick envelope (\ref{Eq: limW_l1})-(\ref{Eq: limW_u2}) is used as a relaxation technique to solve the product of two bounded variables in calculating the apparent power of the DistFlow equations. Engineering constraints are set by the distribution system operator and include node voltage limits, line loadability limits, installed DG capacity limits, reactive power DG limits set by power factors, and investment limits determined by the available budget (\ref{Eq: bgt}).

\subsection{Quality and stability of the solution} \label{Sec:Stability}

Our investment problem formulation described in \mbox{Model \ref{Mod: model}} can be compactly summarizes as a classical two-stage stochastic optimization problem \eqref{m. compact}.

\begin{subequations} \label{m. compact}
\begin{IEEEeqnarray}{r'l}
    z^* = \text{min} & c^Tx + \mathbb{E}_P\left[d^T \mathbf{y} \right]  \label{eq.comp.ob}\\
        \text{s.t.:} &  x \in X  \label{eq.comp.X}\\
        & \mathbf{y} \in \mathbf{Y}(x)  \label{eq.comp.Y}
\end{IEEEeqnarray}
\end{subequations}

The vector $x$ is representing investment decisions at the first stage while the random vector $\mathbf{y}$ represents the operational decisions at the second stage. The objective \eqref{eq.comp.ob} aims to minimize the capital and expected operational costs. At the same time, the budget-limit constrains and power grid operational feasibility constraints are represented by \eqref{eq.comp.X} and \eqref{eq.comp.Y}, respectively. The symbol $\mathbb{E}_P$ is the expected operator over the random parameter distribution $P$. If $P$ represents a continuous distribution vector, this problem is infinite-dimensional, and different approaches have been proposed for solving it.


In rest of this subsection, we describe the metrics to evaluate the investment solution's quality and stability. We will use the sample-average approximation (SAA) method,  \cite{Seljom2019SampleDesign} for approximating the problem \eqref{m. compact}.  

A particular feature of this work is that data is collected from a real grid. Thus, instead of inferring continuous parametric distributions, we use directly observed data in the investment problem addressed here. Still, data can potentially be massive, so we need to find means to reduce the problem's computational complexity. We denote by $N$ to the total number of collected data points, i.e., scenarios. 

In this work, the scenarios are generated using the k-means clustering technique explained in Section \ref{Sec:Kmeans}. To discuss the optimal value limits, we assume the following: (i) the solution has a mean and finite variance. (ii) scenario sampling can be generated in different numbers. (iii) the objective function can be evaluated for the stochastic parameters of the generated outputs. 

\subsubsection{Lower bound estimation}

Using the SAA algorithm, we estimate the value of the lower bound of the DG planning problem's investment solution.  To evaluate the lower bound, we solve $m$ replicas of the two-stage problem \eqref{m. compact} with $n$ scenarios (where $n<N$). We initially generate $m$ sample candidates independently with $n$ scenarios and then solve the approximated sample-based optimization problem \eqref{LB.prob}. Optimal objective of this problem is a lower bound of the original problem \eqref{m. compact}, i.e., $\mathbf{LB}_m(n) \leq z^*$ for any replica $m$.  Because the $n$-drawn scenarios are random, the $\textbf{LB}$ is also random parameter.  

\begin{subequations} \label{LB.prob}
\begin{IEEEeqnarray}{r'll}
    \label{Eq:Two_stage}
    \textbf{LB}_{m}(n) =  \text{min } & c^Tx + \frac{1}{n}\sum_{k=1}^n d^T y_{k}  \\
   \text{s.t.:}  &  x \in X   \\
                &  y_k \in Y_k(x) 
\end{IEEEeqnarray}
\end{subequations}

\subsubsection{Upper bound estimation}

Given a trial (not necessarily optimal) solution for the first stage decision variables denoted by $\hat{x}$, we can compute an upper bound of the original problem \eqref{m. compact} by \eqref{UB_original}, i.e., $z^* \leq \text{UB}_m(\hat{x})$.

\begin{IEEEeqnarray}{r'l}
    \text{UB}_m(\hat{x}) =  c^T\hat{x} + \mathbb{E}_P \Big[ 
    \min_{\mathbf{y}  \in \mathbf{Y}(\hat{x})} & d^T \mathbf{y} 
    \Big]   \label{UB_original}
\end{IEEEeqnarray}

Optimization problem \eqref{UB_original} is scenario-decomposable due to the fixed value of the first-stage decision variables. When the probability distribution function $P$ is discrete, the expected value can be computed exactly compute for each possible random states that can be observed (scenarios). However, if the number of discrete values of the probability distribution $P$ is large or $P$ is continuous, we can approach the upper bound by the \eqref{UB_approx} --  Law of large numbers.

\begin{subequations} \label{UB_approx}
\begin{IEEEeqnarray}{r'l}
    \textbf{UB}_m(\hat{x}, \widehat{N}) =  c^T\hat{x} + \frac{1}{\widehat{N}} \sum_{k=1}^{\widehat{N}} + \;  \text{min } & d^T y_{k}  \\
     \text{s.t.:} &   y_k \in Y_k(\hat{x}) \quad 
\end{IEEEeqnarray}
\end{subequations}

The firs observation is that $\textbf{UB}_m(\hat{x}, \widehat{N})$ is random when $\hat{N}$ random scenarios are drawn. The second observation is that for discrete distributions, as in this paper, the random  $\textbf{UB}_m(\hat{x}, \widehat{N})$ should be approaching to the deterministic  $\text{UB}_m(\hat{x})$ when $\widehat{N} \to N$.   



Finally, we can estimate the optimal solution gap between the lower and upper bounds (\ref{Eq:gap}), that gives statistical information about the stability of the problem. 

\begin{IEEEeqnarray}{rCl}
    \label{Eq:gap}
    \textbf{gap}_{m}(\hat{x}, n, \widehat{N}) &=& \textbf{UB}_m(\hat{x}, \widehat{N}) - \textbf{LB}_{m}(n)
\end{IEEEeqnarray}


\section{Scenario generation and test case} \label{Sec:Test}

There are several methods for generating scenarios from a known probability distribution or a large historical data set. In this section we present the scenario generation technique based on clustering (Section \ref{Sec:Kmeans}). In addition, we present a description of the case study for the computational tests in Section \ref{Sec:Case_test}. 

\subsection{Scenario generation} \label{Sec:Kmeans}

In the stochastic programming model, we analyze four parameters of uncertainty: solar radiation (W/m2), wind speed (m/s), temperature (°C), and active power consumption (W). We use a set of historical data measured with a weather station for weather data and a power meter for active power data. The database has hourly measurements of the uncertainty parameters over one year of recording. The technique used for the generation of scenarios is the k-means clustering technique \cite{Jain2010DataK-means}. The k-means technique is a method to create representative clusters of a data group, whose partitions are given in $k$ clusters. All $k$ clusters have a centroid representing the mean value of the uncertainty parameters contained in that set, minimizing variances within each cluster. 

\begin{figure}[htbp]
    \centering
    \includegraphics[width=\columnwidth]{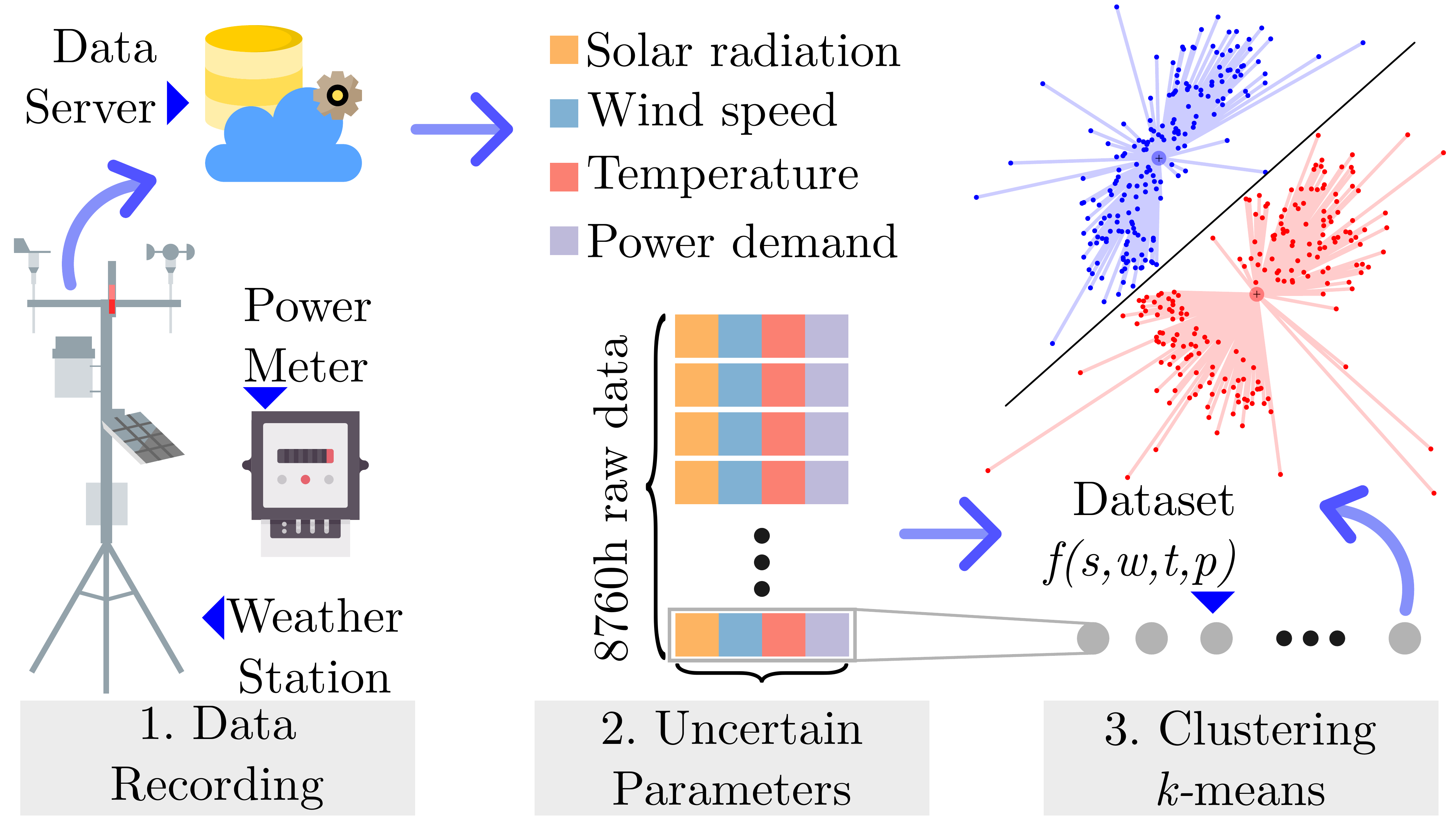}
    \caption{Scenario generation methodology.}
    \label{fig:Scenario}
\end{figure}

The generation of scenarios is done using the historical record of uncertainty parameters (Fig. \ref{fig:Scenario}). Initially, we generate a $k$ number of clusters containing representative data of solar radiation, wind speed, temperature, and power demand. Then we calculate the probability of occurrence of that scenario depending on the cluster's size (amount of data it represents) over the total of registered empirical scenarios. Then, the weather variables are the input to the power production models (\ref{Eq: PV_model})-(\ref{Eq: Wind_model}) of the DG units. The power production model of the PV units depends on solar radiation and ambient temperature, as worked in \cite{Atwa2010OptimalMinimization, Riffonneau2011OptimalBatteries}. The power production model of WT depends only on wind speed. All variables are standardized to fit later on the distribution system. 

\begin{IEEEeqnarray}{rCl}
    P^\text{PV} &=& Y^\text{PV}\left( \frac{G_T}{{G}_{T}^\text{STC}}\right)\left[1-\alpha \left( T_c - T_{c}^\text{STC}\right) \right]
    \label{Eq: PV_model}
    \IEEEeqnarraynumspace\\
    T_c &=& T_a + \frac{G_T}{G_{T}^\text{NOCT}}\left(T_{c}^\text{NOCT} - T_{a}^\text{NOCT} \right)
    \label{Eq: Temp_model}
    \IEEEeqnarraynumspace\\
    P^\text{WT} &=& 
    \begin{cases}
        Y^\text{WT} \frac{v - v_i}{v_r - v_i}, & v_i \leq v < v_r \\
        Y^\text{WT}, & v_r \leq v < v_o\\
        0, & \text{otherwise}
    \end{cases}
    \label{Eq: Wind_model}
\end{IEEEeqnarray}

\nomenclature[M]{$Y_\text{PV}, Y_\text{WT}$}{PV array and wind turbine rated capacity. \hfill [kW]}
\nomenclature[M]{$G_T$}{Solar radiation incident on the PV array \hfill [kW/m$^2$]}
\nomenclature[M]{${G}_{T}^\text{NOCT/STC}$}{Solar radiation at NOCT/STC \hfill [kW/m$^2$]}
\nomenclature[M]{$\alpha$}{Temperature coefficient power \hfill [\%/°C]}
\nomenclature[M]{$T_a, T_c$}{Ambient and PV cell temperature \hfill [°C]}
\nomenclature[M]{$T_{c}^\text{NOCT/STC}$}{PV cell temperature under NOCT/STC \hfill [°C]}
\nomenclature[M]{$v_i, v_r, v_o$}{Cut-in, rated, and cut-off wind turbine speeds \hfill [m/s]}

\subsection{Case study} \label{Sec:Case_test}

The stability analysis of the investment solution is applied to the 34-node distribution system with the topology presented in \cite{Chis1997CapacitorStrategies}. The total installed demand of the system is 5.4 MW with an average power factor of 0.85 in the lag. Historical data was recorded from January 1 to December 31, 2018 with a weather station with an elevation of 36m and 11.02°N - 74.85°W. The two-stage stochastic programming problem was formulated using JuMP v0.21.3 and Gurobi v9.0.1 which provides a solver for the stochastic programming framework. The test machine features OS Name Microsoft Windows Server 2016 Standard, Intel(R) Xeon(R) Gold 6148 CPU @ 2.40GHz, 2394 Mhz, 20 Core(s), 40 Logical Processor(s), Total Physical Memory 256 GB. 

\begin{figure}[htbp]
    \centering
    \includegraphics[width=\columnwidth]{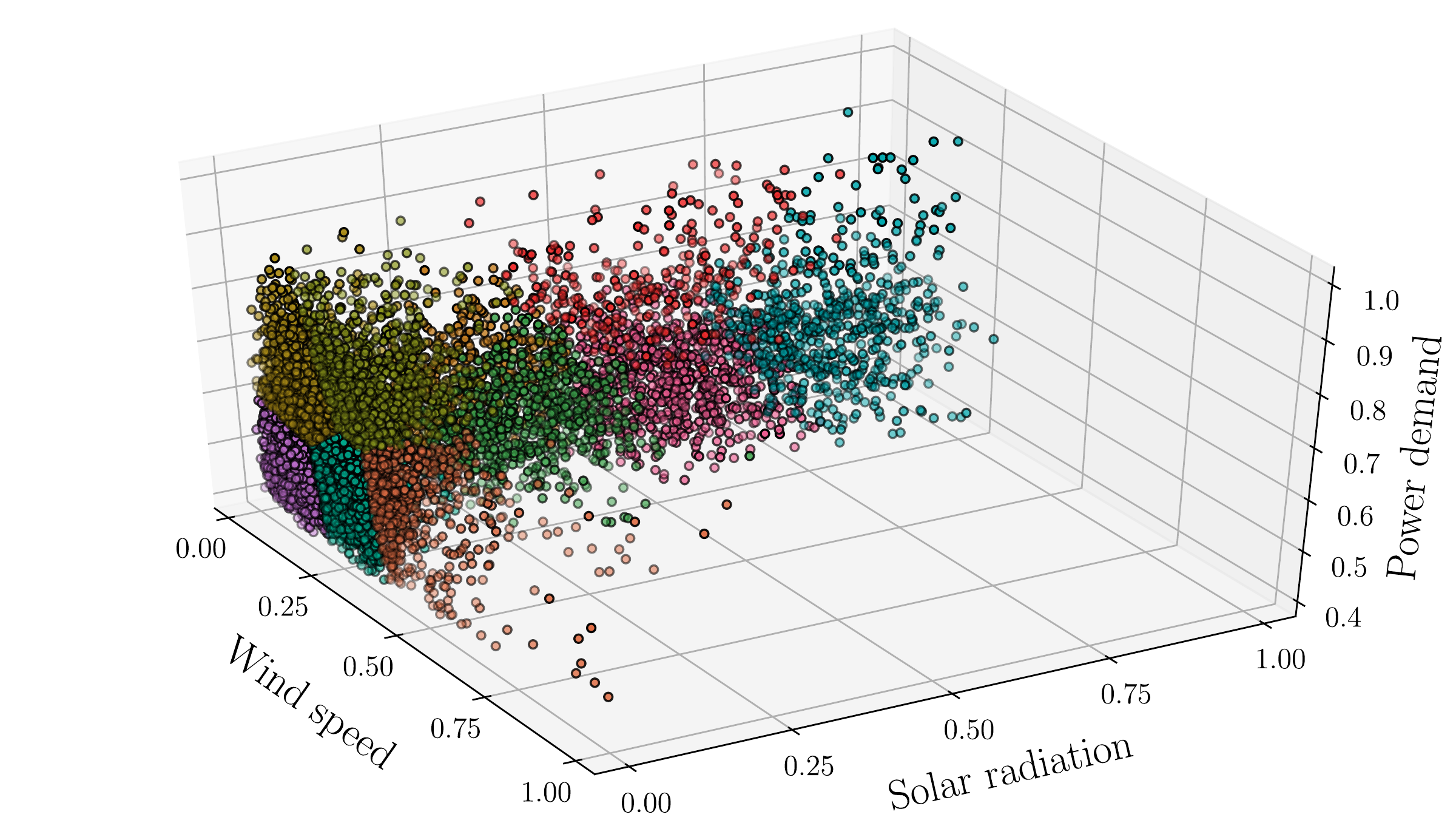}
    \caption{Empirical distribution clustering ($n=10$).}
    \label{fig:Meas_Cluster}
\end{figure}

\section{Results and simulations} \label{Sec:Results}

For the DG planning problem, we apply stability tests for the investment solution with different numbers of generated scenarios. For scenario size $n$, we solve the optimization problem a total of 10 times (replications). The reference value for the solution that we call ground truth is calculated with the maximum number of scenarios that we computationally manage to solve ($n^\prime =5000$). 

\begin{figure}[htbp]
    \centering
    \includegraphics[width=\columnwidth]{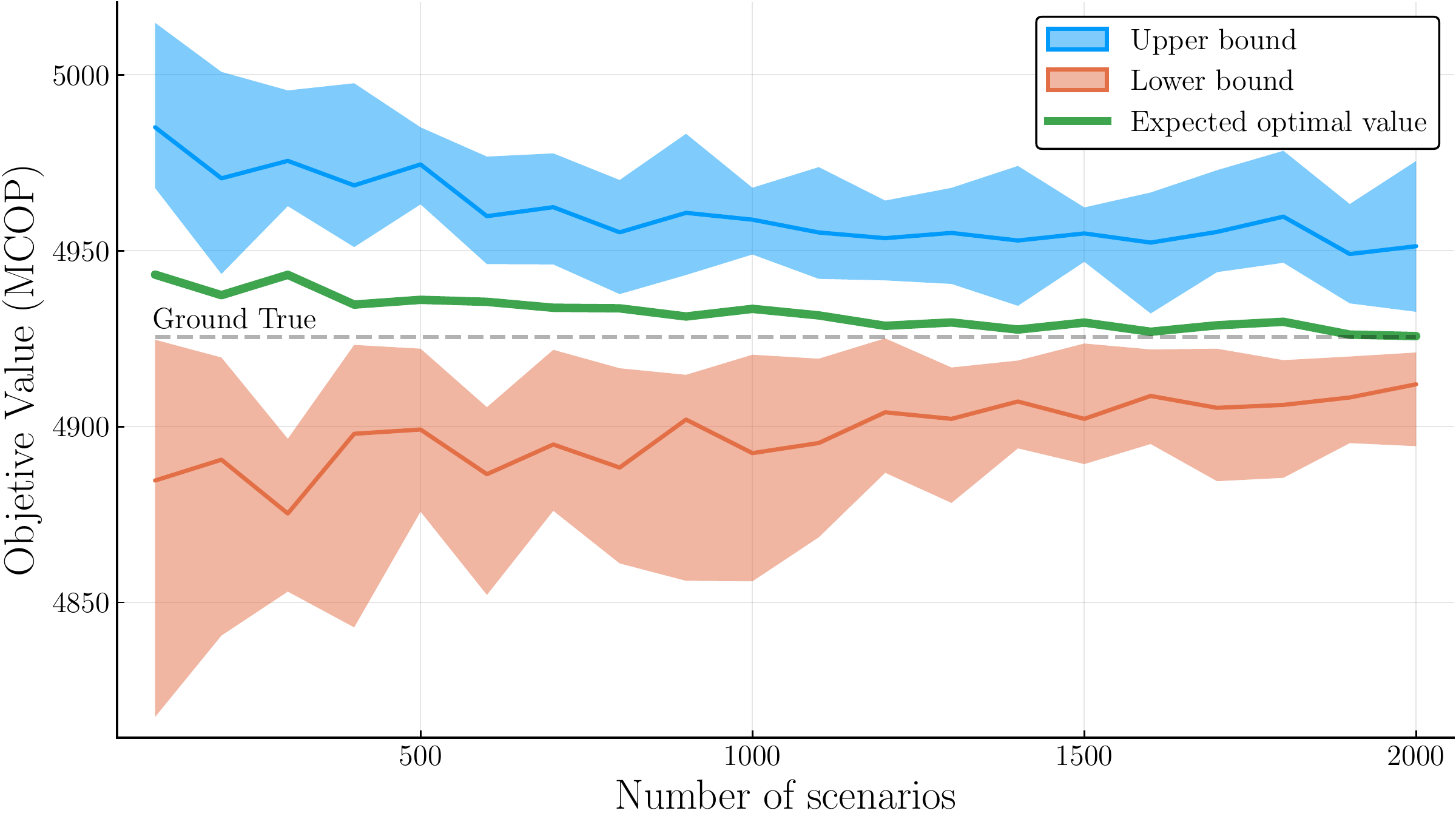}
    \caption{Optimal objective function, estimated lower bound, estimated upper bound.}
    \label{fig:gap}
\end{figure}

\begin{figure}[htbp]
    \centering
    \includegraphics[width=\columnwidth]{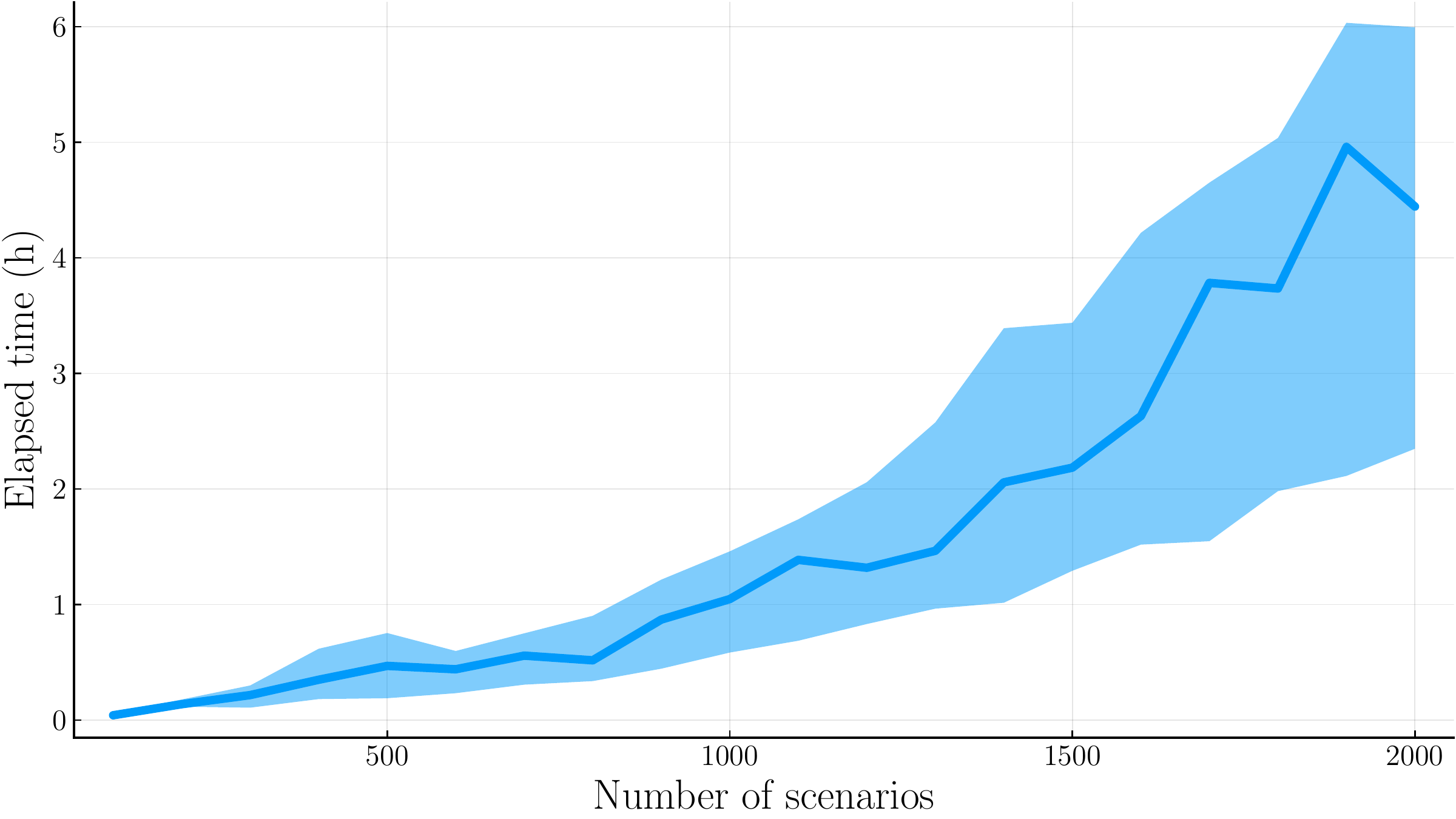}
    \caption{Solution time for the optimization problem.}
    \label{fig:Solve_time}
\end{figure}

Fig. \ref{fig:gap} shows the optimal solution's value solution from (\ref{Eq: of}), the estimated lower boundary, and the estimated upper bound. We can see that the lower bound varies with the number of scenarios generated. The optimal solution's value improves, and the optimality gap size narrows when we increase the generated scenarios' size. This mainly results from the fact that the lower bounds variance decreases as we approach the full empirical distribution. This occurs because the generated scenarios are clustered, and their values may be outside the initial set. The previous problem can be solved with much higher replication values, but it would considerably increase the simulation time (Fig. \ref{fig:Solve_time}). 

The results show that using a few scenarios to solve a stochastic programming problem can lead to substantial errors and sub-optimal solutions. Additionally, the actual probability distribution and stochasticity may not be properly represented in the generated scenarios.

\begin{figure}[htbp]
    \centering
    \includegraphics[width=\columnwidth]{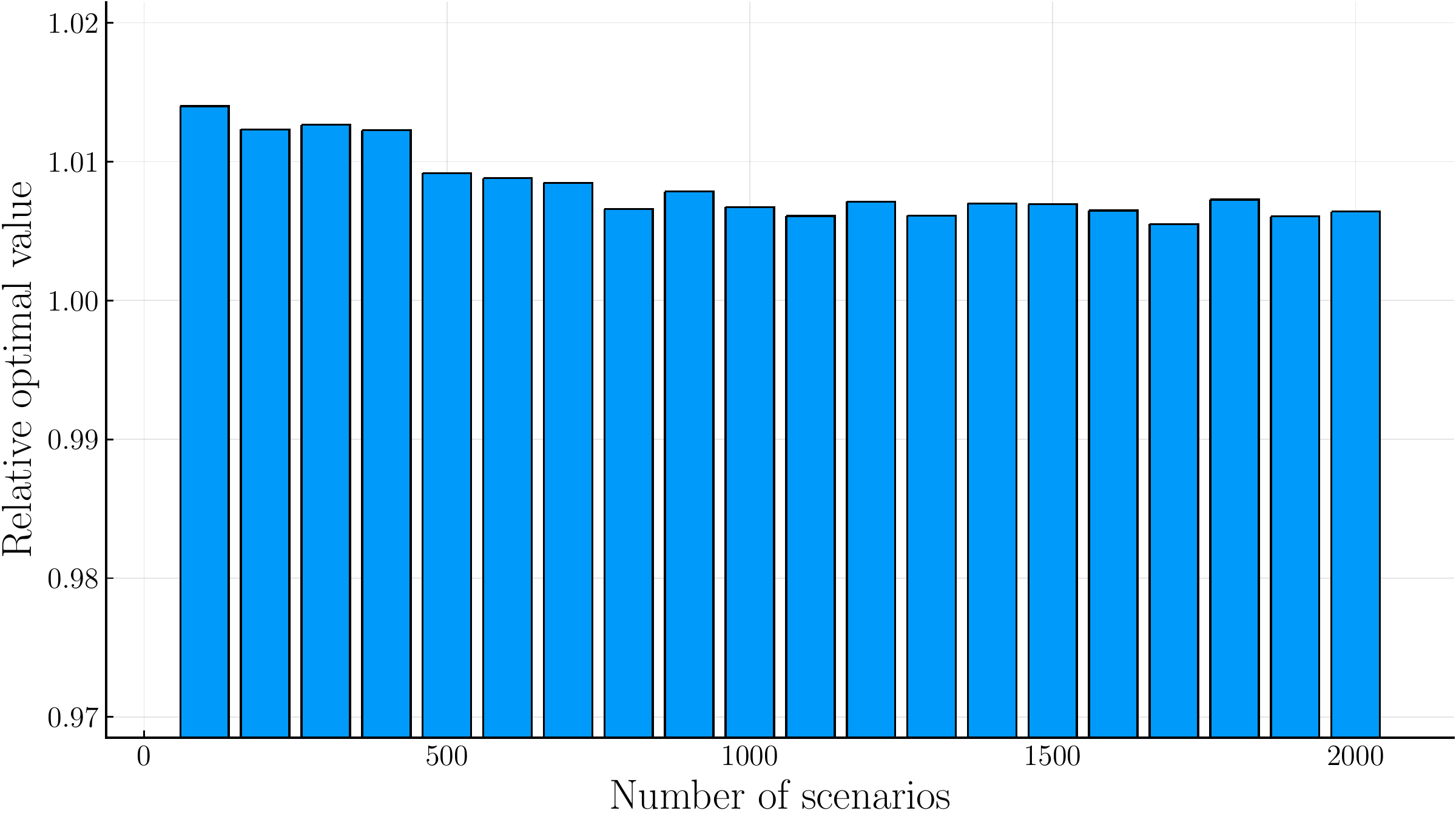}
    \caption{In-sample Stability.}
    \label{fig:InSample}
\end{figure}

\begin{figure}[htbp]
    \centering
    \includegraphics[width=\columnwidth]{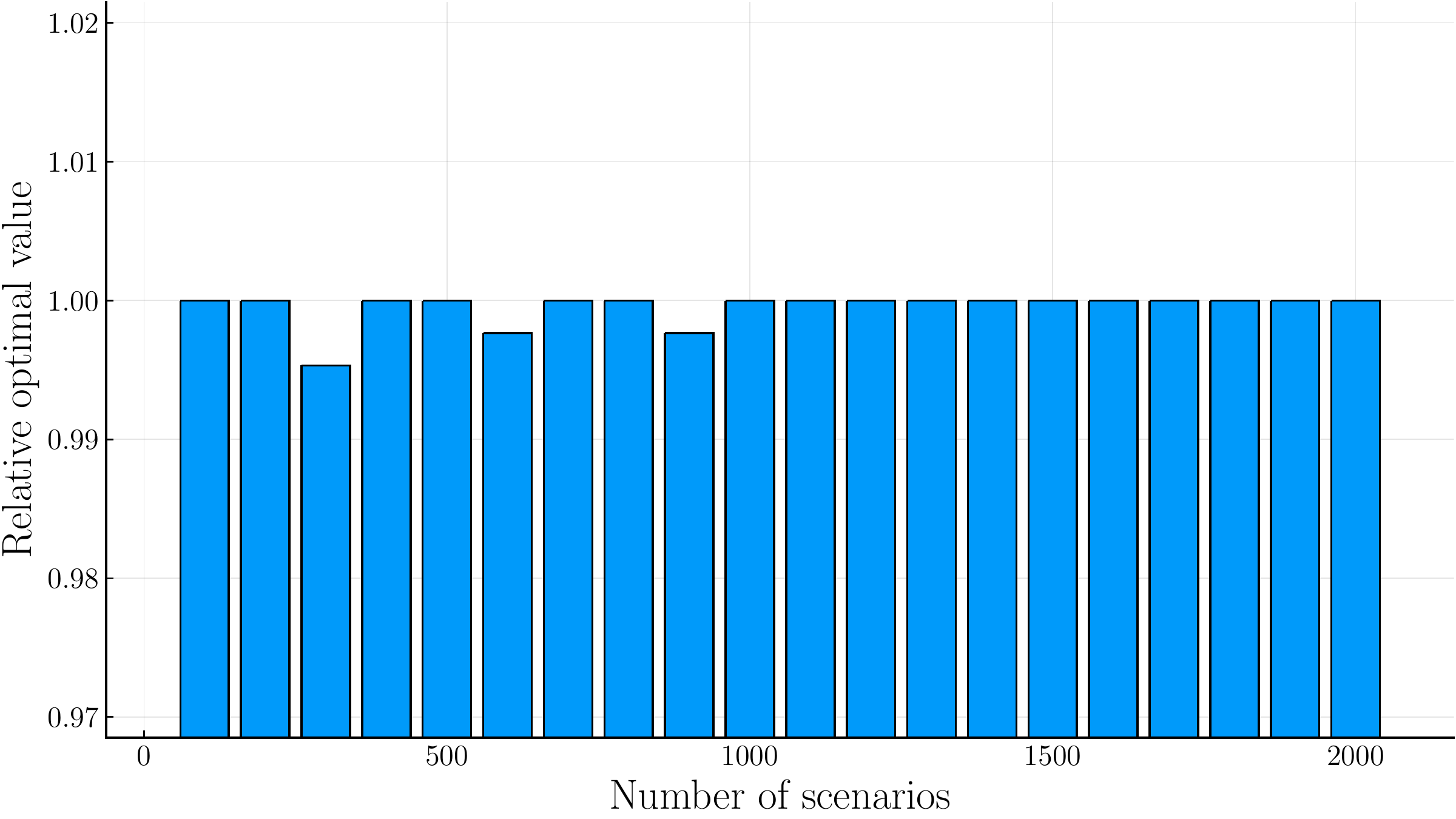}
    \caption{Out-sample Stability.}
    \label{fig:OutSample}
\end{figure}

Fig. \ref{fig:InSample} shows the in-sample stability calculated as the optimal solution's relative value in the $n$ scenario vs. the optimal ground actual value. Also, Fig. \ref{fig:InSample} shows that in-sample stability is improved when we increase the number of scenarios used significantly.  On the other hand, Fig. \ref{fig:OutSample} shows the out-of-sample stability for different numbers of scenarios. The in-sample stability is calculated using the equations and based on our previous notation; the optimal derived values are calculated using the different scenarios (M sets of scenarios with N scenarios each). On the other hand, to calculate out-of-sample stability, we will insert the fixed first-stage solution of each sample m with n size into an optimization problem using the N-scenarios, representing the true distribution. From the figures, we can conclude that high variability in in-sample stability is correlated with high out-of-sample variability. 

\begin{figure}[htbp]
    \centering
    \includegraphics[width=\columnwidth]{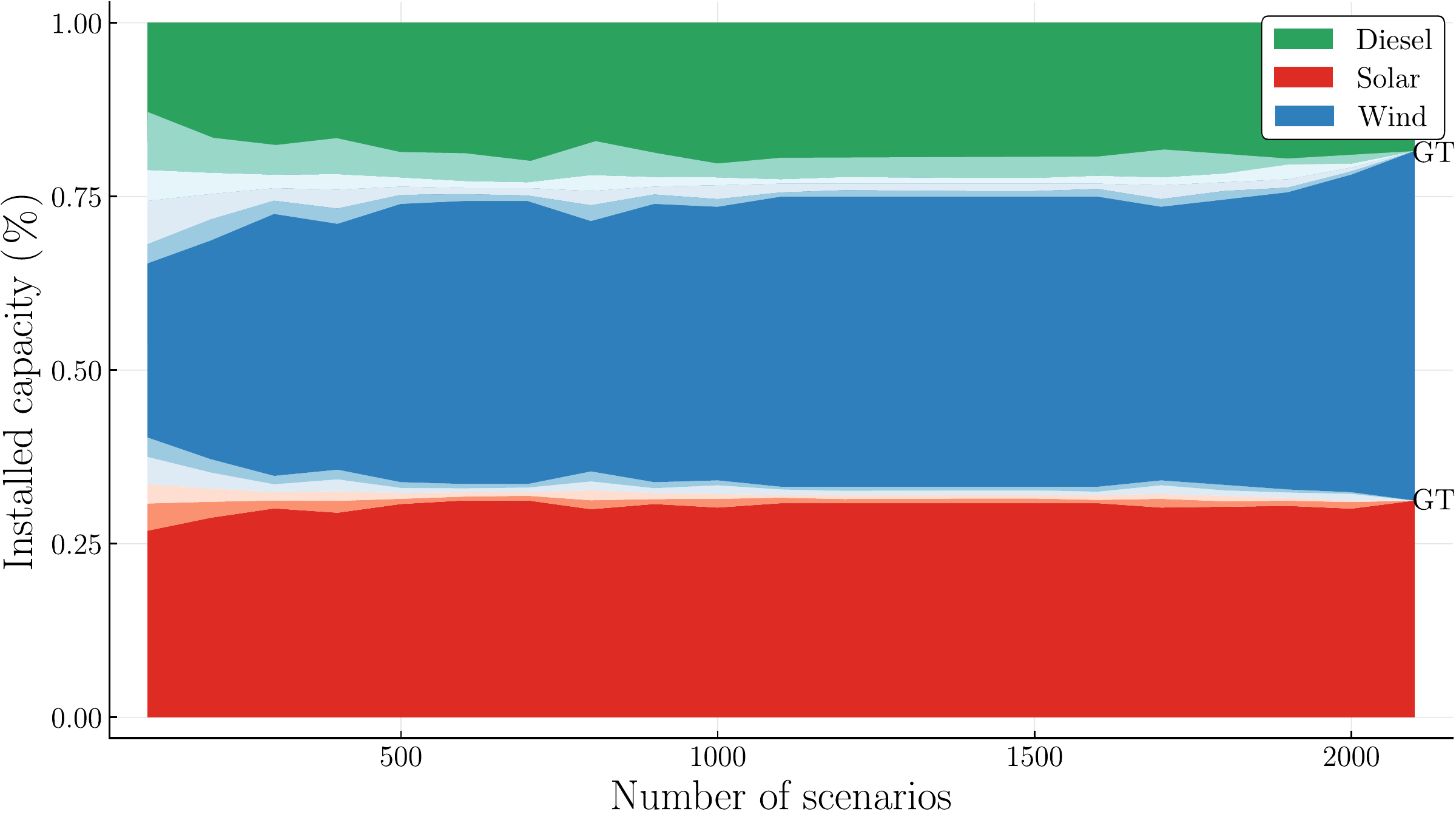}
    \caption{Normalized DG units' installed capacities mix.}
    \label{fig:Stacked_capacity}
\end{figure}

Additionally, we plot the distribution of technologies in a normalized fashion under different numbers of generated scenarios. Fig. \ref{fig:Stacked_capacity} shows the mix of installed capacities when there is no budget constraint. We can see that the installed capacities highly fluctuate when we have a small number of scenarios, while that variability becomes smaller when we have a more significant number of scenarios.  Analogously,

\section{Conclusions} \label{Sec:Conclusions}

This paper applies the SAA technique and stability tests to evaluate the distributed network investment solution's quality using empirical measurements. We show that an investment solution based on a few scenarios can lead to misestimates and deviations from the true solution. On the other hand, the representation of stochasticity and scenarios' use affects the quality of the solution obtained when we have several parameters of uncertainty. From the stability perspective, we can conclude that the solution satisfies the criteria of in-sample and out-of-sample stability when the number of scenarios generated surpasses 1000 data points for this particular distribution grid. In perspective with the stability tests, we can conclude that the model has a good out-of-sample stability performance (the deviations from the optimum relative value are less than 10\% for all scenarios). For the in-sample stability test, we can conclude that for a number of scenarios greater than 500, the deviations from the optimal value are less than 10\%. The SAA algorithm evaluates the solution's quality with the optimality gap using the estimated lower and upper limits. We can conclude that the quality of the solution depends on the number of scenarios used. The optimality gap is reduced to the variability of less than 10\% when the number of scenarios generated is higher than 1000.The optimal of the shared technologies depends on the number of scenarios used; few scenarios lead to a high energy mix variation. A poor representation of the scenarios can lead to an oversize of conventional technologies that derive excess operational costs for the distribution network operator. Finally, we recommend using scenario generation techniques to capture and represent the uncertainty parameters' real distributions adequately. Besides, using as large a number of scenarios as is computationally feasible is highly recommended to find stable and quality solutions to stochastic DG planning.

\bibliographystyle{IEEEtran}
\bibliography{references}

\end{document}

%% file: formulation.tex
\begin{model}[ht]
\caption{Sitting and sizing of distributed generation with non-conventional renewable energies}
\label{Mod: model}

\mbox{\bf Objective:}
\begin{IEEEeqnarray}{C}
    \text{min} \; (\pi^\text{inv} + \pi^\text{OM})
    \label{Eq: of}
\end{IEEEeqnarray}

\mbox{\bf Constraints:}
\begin{IEEEeqnarray}{l}
    \pi^\text{inv} = \smashoperator{\sum_{n, \T}} \left( \pi^{\text{inv}, \T}x^\T_n\right)
    \label{Eq: investment}
    \IEEEeqnarraynumspace\\
    \pi^\text{OM} = \smashoperator{\sum_{\tau}} N_\tau \rho_{\tau}\pi^\text{OM}_{\tau}
    \label{Eq: expected_value_prob}
    \IEEEeqnarraynumspace\\
    \pi^\text{OM}_{\tau} = \pi^\text{loss}_{\tau} + \pi^\text{SS}_{\tau} + \pi^\text{DG}_{\tau}
    \label{Eq: sum_om}
    \IEEEeqnarraynumspace\\
    \pi^\text{loss}_{\tau} = \pi^\text{loss} \smashoperator{\sum_{n,m \in L}} S_{b}R_{n,m}i^{2}_{n,m,\tau}
    \label{Eq: loss_sum}
    \IEEEeqnarraynumspace\\
    \pi^\text{SS}_{\tau} = \pi^\text{SS}_{\tau} S_{b}p^\text{SS}_{\tau}
    \label{Eq: cost_SS}
    \IEEEeqnarraynumspace\\
    \pi^\text{DG}_{\tau} = {S_b}\smashoperator{\sum_{n, \T}} \pi^{\text{OM},\T}p^\T_{n,\tau}
    \label{Eq: DG_cost} \IEEEeqnarraynumspace\\
    \gamma_{\tau}^\text{D}P^\text{D}_{m} {=} \smashoperator{\sum_{n,m \in L}} \left(p^{n,m}_{\tau} {-} p^{m,n}_{\tau} \right) {-} R_{n,m}i^{2}_{n,m,\tau} {+} \sum_\T p^\T_{m,\tau} {+} p^\text{SS}_{m,\tau}
    \label{Eq: active_powerflow}
    \IEEEeqnarraynumspace\\
    \gamma_{\tau}^\text{D}Q^\text{D}_{m} {=} \smashoperator{\sum_{n,m \in L}}\left(q^{n,m}_{\tau} {-} q^{m,n}_{\tau} \right) {-} X_{n,m}i^{2}_{n,m,\tau} {+} \sum_\T q^\T_{m,\tau} {+} q^\text{SS}_{m,\tau}
    \label{Eq: reactive_powerflow}
    \IEEEeqnarraynumspace\\
    {{2} \left( R_{n,m}p^{n,m}_{\tau} {+} X_{n,m}q^{n,m}_{\tau} \right)} {=} v^2_{n,\tau} {+} {\left| Z_{n,m}\right|^2 i^{2}_{n,m,\tau}} {+} v^2_{m,\tau} \hspace{4\jot}
    \label{Eq: current_flow} \\
    w_{n,m,\tau} \ge \underline{V} i^{2}_{n,m,\tau} + v^2_{n,\tau} \underline{I}^{2}_{n,m} - \underline{I}^{2}_{n,m} \underline{V}
    \label{Eq: limW_l1}
    \IEEEeqnarraynumspace \\
    w_{n,m,\tau} \ge \overline{V} i^{2}_{n,m,\tau} + v^2_{n,\tau} \overline{I}^{2}_{n,m} - \overline{I}^{2}_{n,m} \overline{V} 
    \label{Eq: limW_l2}
    \IEEEeqnarraynumspace \\
    w_{n,m,\tau} \le \overline{V} i^{2}_{n,m,\tau} + v^2_{n,\tau} \underline{I}^{2}_{n,m} - \overline{V} \underline{I}^{2}_{n,m}
    \label{Eq: limW_u1}
    \IEEEeqnarraynumspace \\
    w_{n,m,\tau} \le v^2_{n,\tau} \overline{I}^{2}_{n,m} + \underline{V} i^{2}_{n,m,\tau} - \underline{I}^{2}_{n,m} \overline{I}^{2}_{n,m}
    \label{Eq: limW_u2}\\
    \underline{V}^2 \le v^2_{n,\tau} \le \overline{V}^2
    \label{Eq: volt_limit}
    \IEEEeqnarraynumspace\\
    i^{2}_{n,m,\tau} \le \overline{I}^{2}_{n,m}
    \label{Eq: current_limit}
    \IEEEeqnarraynumspace\\
    \overline{P}^\text{PV}x^\text{PV}_n + \overline{P}^\text{WT}x^\text{WT}_n + \overline{P}^\text{CG}x^\text{CG}_n    \le \overline{P}_n
    \label{Eq: dg_node_limits}
    \IEEEeqnarraynumspace\\
    0 \le p^{\T}_{n,\tau} \le \gamma^{\T}_{\tau}
    \overline{P}^{\T}x^{\T}_n\beta^\T_n
    \label{Eq: solar_limits}
    \IEEEeqnarraynumspace\\
    \lambda^{\T,+} p^{\T}_{n,\tau} \le q^{\T}_{n,\tau} \le \lambda^{\T,-} p^{\T}_{n,\tau}
    \label{Eq: PV_pf_lim}
    \IEEEeqnarraynumspace\\
    \lambda^{\T,\text{+/-}} = \mp \tan(\cos^{-1}(\text{pf}^\text{+/-}))
    \label{Eq: pf_lead_eq}
    \IEEEeqnarraynumspace\\
    \pi^\text{inv} \le \Pi^\text{bgt}
    \label{Eq: bgt}
    \IEEEeqnarraynumspace
    \vspace{-4\jot}
\end{IEEEeqnarray}
\nomenclature[I]{$n {\in} N$}{Index/set of power nodes}
\nomenclature[I]{$t {\in} T$}{Index/set of time blocks}
\nomenclature[I]{$\omega {\in} \Omega$}{Index/set of scenarios}
\nomenclature[I]{$(n,m) {\in} L$}{Index/set of lines}
\nomenclature[P]{$S_{b}$}{Apparent power base \hfill [kVA]}
\nomenclature[P]{$R_{n,m}$}{Resistance line from bus n to bus m \hfill [p.u.]}
\nomenclature[P]{$Z_{n,m}$}{Impedance of branch $n$, $m$ \hfill [p.u.]}
\nomenclature[P]{$\overline{P_n}$}{Maximum installable active power at $n$ \hfill [p.u.]}
\nomenclature[P]{$\beta^\T_n$}{Binary parameter for buses available to install DG}
\nomenclature[P]{$\lambda^{\text{+/-},\T}$}{Technologie's lead/lagging power factor \hfill [p.u.]}
\nomenclature[P]{$\text{pf}^{\text{+/-},\T}$}{Technologies' lead/lagging power factor angle \hfill [rad]}
\nomenclature[I]{$\T$}{Set of installable technologies $\text{\{PV, WT, CG\}}$}
\nomenclature[P]{$X_{n,m}$}{Reactance line from bus n to bus m \hfill [p.u.]}
\nomenclature[V]{$\text{Variables are defined within the text in section \ref{Sec:Two_Stage}.}$}{}
\end{model}